\documentclass[onecoloumn]{revtex4}

\usepackage{graphicx}
\usepackage{dcolumn}
\usepackage{bm}
\usepackage{amssymb}
\usepackage{amsmath}

\begin{document}

\title{Velocity Measurements in Some Classes of Alternative Gravity Theories}

\author{Sumanta Chakraborty}

\affiliation{Department Of Physics,\\
 Rajabajar Science College, Calcutta University,
  \\92 A.P.C road, Kolkata-700009,\\
India \footnote{sumantac.physics@gmail.com}}

\begin{abstract}
The general misconception regarding velocity
measurements of a test particle as it approaches black hole is
addressed by introducing generalized observer set. For a general
static spherically symmetric metric applicable to both Einstein
and alternative gravities as well as for some well known
solutions in alternative gravity theories, we find that velocity
of the test particle do not approach that of light at event
horizon by considering ingoing observers and test particles. 
\end{abstract}



\maketitle


\section{Introduction}\label{vi}

The radial motion of a test particle falling in a black hole is
one of the key issues in general relativity. The infalling motion
has been studied specifically for Schwarzschild black hole by
several authors (\cite{lan71},\cite{Wald},\cite{Bergmann},\cite{Moller}). All of them reached the same conclusion that
velocity of the infalling particle approaches that of light near
the event horizon, which for the Schwarzschild case is at $r=2M$,
where $M$ is the mass of the black hole. The observers, called
static observers, are at rest with respect to the mass creating
the gravitational field. They are actually the world lines on the
hypersurface of orthogonal killing vector field for the metric
describing the gravitational field. However there exists a common
misconception that particle approaches the speed of light as it
moves to the black hole horizon for all observers, but not as a
limiting procedure for a static observer at $r$ as $r \rightarrow
2M$. However if we assume that the particle approaches the event
horizon at the speed of light for a static observer, as we have
defined it earlier, then simple velocity composition law tells
that it should approach the speed of light for all local observers
as space time is locally Minkowskian.

So we have to modify our notion of velocity for a test particle
near a black hole for a static observer which was done for
Schwarzschild black hole (\cite{Crawford},\cite{jan77}). The notion of observer is implemented and used in various co-ordinate frames by several
authors (\cite{bol11},\cite{ell85},\cite{bol06}).

However recently a progress has been made in obtaining trajectory around a general spherically symmetric non-rotating black hole by
choosing a general metric ansatz \cite{cha11},

\begin{equation}\label{i1}
ds^{2}=-f(r)dt^{2}+\frac{dr^{2}}{f(r)}+r^{2}d\Omega ^{2}
\end{equation}

For this general case we find the velocity of the test particle
with respect to a static observer ($r=$constant) to be a function
of $f(r)$. While for the case of a general observer such that both
the observer and the test particle moves along geodesic in $\theta
=\frac{\pi}{2}$ plane then the velocity of the test particle with
respect to the observer to our surprise, do not depend on the
choice of the function $f(r)$ provided the particle has high
energy which is the most common case for astrophysical bodies,
however it depends on the angular momenta which was absent in
earlier works \cite{Crawford}. Then we have used some classes of
spherically symmetric solutions in  alternative gravity theories
to find the relative velocity of a test particle with respect to an
observer. We have discussed spherically symmetric solution in
string inspired dilaton model \cite{Garfinkle}, and calculate
motion of a test particle in this spacetime. Secondly we have
considered a spherically symmetric solution in quadratic gravity
obtained in a recent paper \cite{Yunes} to discuss the velocity
profile of an object. Finally we have discussed motion in
spherically symmetric solutions in
Einstein-Maxwell-Gauss-Bonnet(EMGB) theory and vacuum solution in
$F(R)$ gravity. Throughout the paper we shall use natural unit
such that $G=c=1$.

This paper is organized as follows, in section ($\ref{vc}$) we introduce the general idea of observer and co-ordinate frames which we shall use
throughout this work. In section ($\ref{vs}$) we discuss the motion in spherical symmetric space-time for the general choice of metric
as presented in equation ($\ref{i1}$). In the next section we discuss some classes of alternative gravity theories. The paper ends with a
short discussion on the
results obtained.

\section{Co-ordinate System, Reference Frames and Observers}\label{vc}

The mathematical beauty of general relativity is the freedom of
choice of coordinates in the description of physical phenomenon.
We could choose any co-ordinate system as we wish, this choice
might be taken in favor of the symmetry involved in the problem.
Also the co-ordinates are not sufficient we need reference frame
as well. However the co-ordinate system and reference frames are
not independent, for example in one reference frame one set of
co-ordinates may be important while it could change in other
reference frame. However in literature \cite{Bergmann} it is
often seen that co-ordinate system and reference frames are used
interchangeably. However in our discussion we find the use of
"reference frame" and "co-ordinate system" to be distinct. By
reference frame we shall mean a set of observers to take
measurements, for example the set of all observers moving in a
time like geodesic form a reference frame, whereas co-ordinate
system refer to numbers specified over the whole space time
manifold.

 In special relativity an infinite lattice work of sticks and clocks \cite{Wheeler} suffice to define a unique reference frame.
 However in general relativity we cannot have such rigid framework since the space time is Minkowskian only locally, so we replace this rigid system
 by a fluid \cite{Moller}. In a strictly mathematical sense the set of observers represents a set of future pointing time like congruence,
 which is a three parameter family of curves $x^{\mu}(\lambda ,y^{i})$, where $\lambda$ is an affine parameter defined over the path, and $y^{i}$
 labels the spatial parts of the curve.

 Observer in general theory is very local and it is a material particle parameterized by proper time. An observer field i.e. its velocity field
 $u$ on the manifold $M$ is stationary provided there exist a smooth function $f$ greater than $0$, such that $fu=\xi$ is a killing vector field,
 so the lie derivative of the metric with respect to the vector field $\xi$ vanishes
 (i.e. $L_{\xi}g_{\mu \nu}=0$).

 There is a natural way for an $u$-observer to define the speed of any particle with four velocity $t^{\mu}$ as it passes an event $p\in M$,
 then the observer measure the square of the speed at event $p$ to yield \cite{Crawford},

\begin{equation}\label{i4}
v^{2}=\frac{(g_{\mu \nu}+u_{\mu}u_{\nu})t^{\mu}t^{\nu}}{(u_{\alpha}t^{\alpha})^{2}}
\end{equation}

Then we have $g_{\mu \nu}t^{\mu}t^{\nu}=-1$ and as well as $u_{\mu}u_{\nu}t^{\mu}t^{\nu}=(u_{\alpha}t^{\alpha})^{2}$.
Thus the above relation can be simplified to yield,

\begin{equation}\label{i3}
v^{2}=1-\frac{1}{(u^{\mu}t_{\mu})^{2}}
\end{equation}

Note that the two velocities $u^{\mu}$ and $t^{\mu}$ are time like
as observer and the test particle are both time like.

\section{Motion in a General Spherically Symmetric Space Time}\label{vs}

\subsection{Test Particle Geodesic}

We shall assume that our test particle is confined to a plane
which is generally chosen as $\theta=\pi/2$ for calculational
simplicity and as well as we have spherical symmetry so if we
discuss the situation for some specified $\theta$ plane then it
would be the same for all. Thus this no longer represent a
radially ingoing particle but a more generalized case where the
particle has two variables to specify namely, ($r,\phi$). The
motion is determined by the Euler equations corresponding to the
lagrangian formed as $2L=g_{\mu \nu}\dot{x}_{\mu}\dot{x}_{\nu}$,
Which has the following explicit form (using equation
($\ref{i1}$)),

\begin{equation}\label{v1}
 2L=-f(r)\dot{t}^{2}+\frac{1}{f(r)}\dot{r}^{2}+r^{2}\dot{\phi}^{2}
\end{equation}

where dot denotes differentiation with respect to proper time of
the particle. This equation can be written in terms of the
particle proper time and then along the orbit we have $2L=-1$.
This finally leads to,

\begin{equation}\label{v2}
 d\tau ^{2}=f(r)dt^{2}-\frac{1}{f(r)}dr^{2}-r^{2}d\phi ^{2}=f(r)dt^{2}[1-v^{2}]
\end{equation}

where

\begin{equation}\label{v3}
 v^{2}=\frac{1}{f(r)^{2}}\left(\frac{dr}{dt}\right)^{2}+\frac{r^{2}}{f(r)}\left(\frac{d\phi}{dt}\right)^{2}
\end{equation}

This is the velocity of the particle with respect to a static observer (r=constant) as illustrated by plugging $u^{\mu}=(1,0,0,0)$ in
equation (\ref{i4}); i.e. the particle moves through a distance $\frac{1}{\sqrt{f}}\sqrt{dr^{2}+r^{2}fd\phi ^{2}}$ in a proper time given by
$\sqrt{f}dt$, where from now on we shall use simply $f$ for $f(r)$ due to notational simplicity.

 Since the lagrangian as given in ($\ref{v1}$) do not contain $t$ explicitly we have a constant of motion which is nothing but the
energy of the particle and it is given by,

\begin{equation}\label{v4}
 -\frac{\partial L}{\partial \dot{t}}=f\dot{t}=E
\end{equation}

This constant of motion actually originates from the killing
vector field $\frac{\partial}{\partial t}$, this can be phrased
as, if the 4-velocity of the particle $t^{a}$ is a geodesic, then
we have $\triangledown _{t}t=0$. From ($\ref{v2}$) and
($\ref{v4}$) we have obtained,

\begin{equation}\label{v5}
 v=\sqrt{1-\frac{f}{E^{2}}}
\end{equation}

Also the energy can be  determined from the initial value of
radius and velocity using ($\ref{v5}$) such that,
$E^{2}=\frac{f(R)}{1-v_{0}^{2}}$. Where $R$ is the initial radial
co-ordinate and $v_{0}$ is the initial velocity.

 We have another constant of motion in this case which corresponds to the angular momentum of the particle and could be given by,

\begin{equation}\label{v6}
 \dot{\phi}=\frac{L}{r^{2}}
\end{equation}

Thus finally the velocity in proper frame on the plane $\theta =\frac{\pi}{2}$ is given by,

\begin{equation}\label{v7}
 \left(\frac{dr}{d\tau}\right)^{2}=\left(\frac{dr}{dt}\right)^{2}\left(\frac{dt}{d\tau}\right)^{2}=E^{2}-V^{2}
\end{equation}

where $V^{2}=f\left[1+\frac{r^{2}}{f^{2}}E^{2}\left(\frac{d\phi}{dt}\right)^{2}\right]=f\left[1+\frac{L^{2}}{r^{2}}\right]$.

Thus the 4-velocity components for the geodesic particle specified by energy and angular momentum is given by,

\begin{equation}\label{v8}
 t^{\mu}=\left(\frac{E}{f},\sqrt{E^{2}-V^{2}},0,\frac{L}{r^{2}}\right)
\end{equation}

written in terms of the constants of motion $E$ and $L$. As a check we can use the identity $t_{\mu}t^{\mu}=-1$.
Thus equation (\ref{v8}) represents the four velocity of a test particle in a space-time metric given by equation (\ref{i1}).

\subsection{Static Limit}

In some cases the velocity is measured in terms of proper time, as determined by clocks synchronized along trajectory of the
particle. The velocity in case of radial particle is given by \cite{lan71},

\begin{equation}\label{v9}
 v^{2}=\left(g_{00}+g_{01}\frac{dx^{1}}{dx^{0}}\right)^{-2}\left(g_{01}^{2}-g_{00}g_{11}\right)\left(\frac{dx^{1}}{dx^{0}}\right)^{2}
\end{equation}

When we generalize this result to our case where we have three
co-ordinates $x^{0},x^{1}$ and $x^{3}$ (since $x^{2}=\theta
=constant$), then velocity expression generalizes to,

\begin{equation}\label{v10}
 v^{2}=\frac{\left(g_{10}^{2}-g_{00}g_{11}\right)\left(\frac{dx^{1}}{dx^{0}}\right)^{2}+
\left(g_{30}^{2}-g_{00}g_{33}\right)\left(\frac{dx^{3}}{dx^{0}}\right)^{2}+
2\left(g_{10}g_{30}-g_{13}g_{00}\right)\frac{dx^{1}}{dx^{0}}\frac{dx^{3}}{dx^{0}}}
{\left(g_{00}+g_{10}\frac{dx^{1}}{dx^{0}}+g_{30}\frac{dx^{3}}{dx^{0}}\right)^{2}}
\end{equation}

Note that if we let $\frac{dx^{3}}{dx^{0}}$ to be zero, then it
reduces to equation ($\ref{v9}$). In our case keeping the non zero
terms we obtain,

\begin{equation}\label{v11}
 v^{2}=\frac{1}{f^{2}}\left( \frac{dr}{dt} \right)^{2}+\frac{r^{2}}{f}\left( \frac{d\phi}{dt} \right)^{2}
\end{equation}

which is completely identical to ($\ref{v3}$). This definition has
co-ordinate invariance. The 4 velocity has components
$u_{\mu}=(-g_{00})^{-1/2}g_{\mu 0}$ and that for the particle
reduces to
$t^{\mu}=(\frac{dx^{0}}{d\tau},\frac{dx^{1}}{d\tau},0,\frac{dx^{3}}{d\tau})$.
Thus using equation ($\ref{i3}$) we obtain the same equation as
($\ref{v11}$).

From equation ($\ref{v5}$) we see that as $f(r)=0$, the velocity
is equal to 1. Hence for static observers $v$ approaches the speed
of light at the event horizon and they predict faster than light
speed inside event horizon.

 It might seem at first sight that this result has nothing to do with $f(r)=0$ but is connected to the co-ordinate system.
 However it has nothing to do with co-ordinate system but with the observer. So we should generalize our observer set.

 Also no observer can be at rest at $r=2M$ except photon, with respect to photon all particle traverse at speed of light.
 To get a clear view we discuss the acceleration of a static observer in the field of the gravitating body.
 The acceleration is necessary as in general relativity an observer at rest is not geodesic and is accelerated.

 The four acceleration field is defined as,

\begin{equation}\label{v12}
 a^{\eta}=u^{\eta}_{;\mu}u^{\mu}=\left(u^{\eta}_{,\mu}+u^{\alpha}\Gamma ^{\eta}_{\alpha \mu}\right)
\end{equation}

The only non zero component is given by using the definition of four velocities for static observers , $u_{\mu}=\frac{g_{\mu 0}}{\sqrt{-g_{00}}}$
to yield,

\begin{equation}\label{v13}
 a^{1}=\frac{1}{2}\frac{df}{dr}
\end{equation}

So acceleration depends on the function $f(r)$.
\subsection{Ingoing Observers}


\begin{figure}
\includegraphics[height=3.5in, width=3.5in]{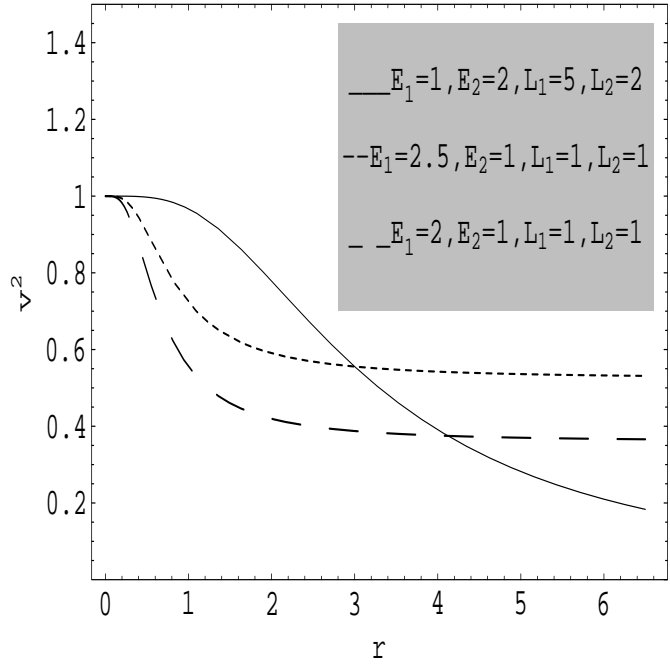}

\caption{The figure shows variation of $v^{2}$ with radial co-ordinate r for different choice of $E_{1}$, $E_{2}$, $L_{1}$ and $L_{2}$.\label{fig1}}

\end{figure}
\begin{figure}
\includegraphics[height=3.5in, width=3.5in]{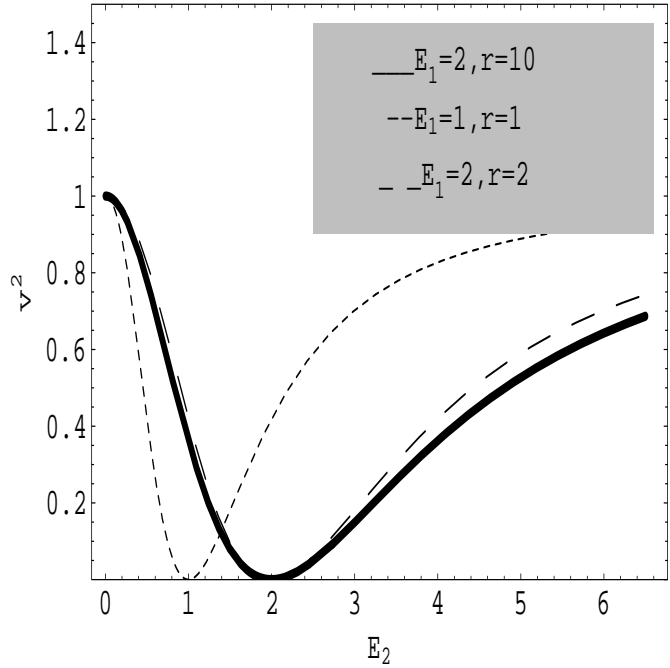}

\caption{The figure shows variation of $v^{2}$ with
test particle energy for different observer energy and radial
distance.\label{fig2}}

\end{figure}
\begin{figure}
\includegraphics[height=3.5in, width=3.5in]{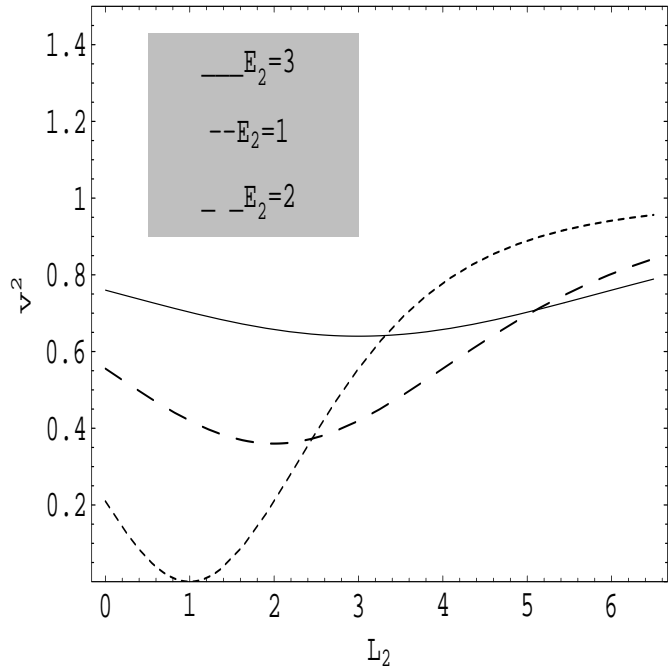}

\caption{The figure shows variation of $v^{2}$ with
test particle angular momentum for different choices of test
particle energy.\label{fig3}}

\end{figure}
\begin{figure}
\includegraphics[height=3.5in, width=3.5in]{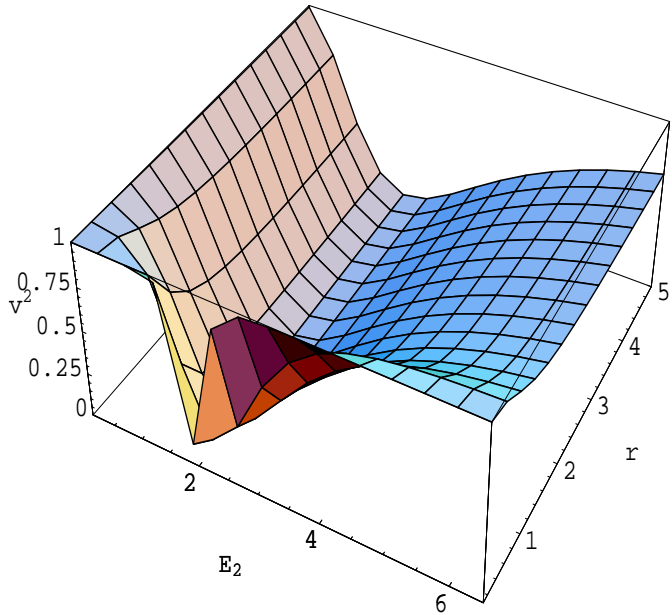}

\caption{The figure shows variation of $v^{2}$ with $E_{2}$ and r.\label{fig4}}

\end{figure}

We consider motion of two particles such that the four velocities are given by,

\begin{eqnarray}\label{v14}
\left.\begin{array}{c}
t^{\mu}=\left( \frac{E_{1}}{f}, \sqrt{E^{2}_{1}-V^{2}_{1}}, 0, \frac{L^{2}_{1}}{r^{2}} \right)\\
u^{\nu}=\left( \frac{E_{2}}{f}, \sqrt{E^{2}_{2}-V^{2}_{1}}, 0, \frac{L^{2}_{2}}{r^{2}} \right)
\end{array}\right\}
\end{eqnarray}

Hence we obtain the following result,

\begin{equation}\label{v15}
t^{\mu}u_{\mu}=g_{\mu \nu}t^{\mu}u^{\nu}=-\frac{E_{1}E_{2}}{f}+
\frac{\sqrt{\left( E^{2}_{1}-V^{2}_{1} \right) \left( E^{2}_{2}-V^{2}_{1} \right)}}{f}+\frac{L_{1}L_{2}}{r^{2}}
\end{equation}

Thus we obtain,

\begin{equation}\label{v16}
\left( t^{\mu}u_{\mu} \right)^{2}=\left( \frac{E_{1}E_{2}}{f} \right)^{2}
\left[ 1-\frac{fL_{1}L_{2}}{r^{2}E_{1}E_{2}}-\sqrt{\left( 1-fa_{1} \right)\left( 1-fa_{2} \right)} \right]^{2}
\end{equation}

Where, $a_{1}=\frac{1}{E^{2}_{1}}\left( 1+\frac{L^{2}_{1}}{r^{2}} \right)$ and
$a_{2}=\frac{1}{E^{2}_{2}}\left( 1+\frac{L^{2}_{2}}{r^{2}} \right)$.

Simplifying and rearranging terms we have obtained that,

\begin{equation}\label{v17}
 \left( t^{\mu}u_{\mu} \right)^{2}=E^{2}_{1}E^{2}_{2}
\left[\frac{1}{2}\left( a_{1}+a_{2} \right)-\frac{L_{1}L_{2}}{r^{2}E_{1}E_{2}} \right]^{2}
\end{equation}

We know that the relative velocity could be given by, $v^{2}=1-\frac{1}{\left(u^{\mu}t_{\mu}\right)^{2}}$. Thus using equation ($\ref{v17}$) and assuming that energy of both the particle and the observer are high enough or the distance is large enough we ultimately arrive at,

\begin{equation}\label{v18}
v^{2}=1-\frac{4}{E^{2}_{1}E^{2}_{2}\left[ \left( \frac{1}{E^{2}_{1}}+\frac{1}{E^{2}_{2}} \right)+
\frac{1}{r^{2}}\left( \frac{L_{1}}{E_{1}}-\frac{L_{2}}{E_{2}} \right)^{2} \right]^{2}}
\end{equation}

Note that as $r\rightarrow 0$ the velocity approaches that of
light i.e. $v=1$. However if the particle and the observer has the
same impact parameter i.e.
$\frac{L_{1}}{E_{1}}=\frac{L_{2}}{E_{2}}$ then even if
$r\rightarrow 0$ the velocity does not approach $1$, which is a
very interesting result. Also at short distance the velocities and
hence energies are very high so $a_{1}$ and $a_{2}$ are small
quantities, however at large distance not $a_{1}$ and $a_{2}$ but
$f(r)$ become smaller and thus as they appear in product form in
the velocity expression it holds good for all $r$. Thus we can say
that equation ($\ref{v18}$) is a general result. This result is
valid in spherically symmetric solutions for Einstein gravity like
the Schwarzschild and Reissner-Nordstr\"{o}m solutions but also
for the Einstein-Maxwell-Gauss Bonnet theory. There exists two
additional well known spherically symmetric solution but they do
not have the form used. So we shall consider them in the next two
sections.

 Note from figure-$\ref{fig1}$, as radial co-ordinate of the particle is decreased the velocity remain less than speed of light.
 As $r\rightarrow 0$ the velocity also approaches $1$ in our system of units, which is justified and shows the actual motion that happen as the
 particle moves within the event horizon. It is also clear that with increase of the energy of the particle the velocity increases and it also
 increases with increasing the angular momenta.

 From figure-$\ref{fig2}$ we see that as energy of the particle is increased we get a interesting behavior, at first it decreases and become zero,
 then it again increases. Thus here the combined quantity in the denominator becomes 4. This happens when $E_{1}$ coincides with $E_{2}$
 (see the figure), as we have chosen $L_{1}=L_{2}$ (see equation $\ref{v18}$). However changing the radius has a very small effect on velocity profile.

 From figure-$\ref{fig3}$ we find that velocity varies with angular momentum in some what the same manner as it does with energy.
 However by proper choice of $E_{2}=E_{1}$ the velocity can be made zero when $L_{2}=E_{2}$ as we have chosen other parameters such that $L_{1}=E_{1}$,
 since under this condition the denominator in equation ($\ref{v18}$) become $4$. As well as we can eliminate that zero by changing $E_{2}$.

 Figure-$\ref{fig4}$ shows the variation of velocity both with radial co-ordinate and the energy of the particle. This graph merely shows combined effects of
 varying radius and energy as we have illustrated in earlier graphs.

\section{Motion for Some Classes of Alternative Gravity Theories}\label{vs2}

Current theoretical cosmology has two fundamental problems, namely inflation and late time acceleration of the universe. The usual scenarios used to explain both these accelerating cosmology epochs are to develop acceptable dark energy model, such as: scalar, spinor, cosmological constant and higher dimensions. Even if such a scenario seems to be partially succesful it is hindered by the coupling with usual matter, compatibility with standard elementary particle theories.

However another natural choice is the classical generalization of general relativity, called modified gravity or alternative gravity theory (\cite{cald03}, \cite{noj03},\cite{noj07},\cite{noj11}). Thus a gravitational alternative to explain inflation and dark energy seems very reasonable on the ground of the expectation that general relativity is just an approaximation that is valid at small curvature. A sector of modified gravity containing the gravitational terms relevent at high energy produced the inflationary epoch. During evolution curvature decreases and general relativity describes to an good approaximation the intermediate universe. With a furthur decrease of curvature as sub-dominant terms grow we see a transition from deceleration to cosmic acceleration. There exists traditional $F(R)$, string inspired models, scalar tensor theory, Gauss-Bonnet theory and some other models. In the next subsections we shall discuss motion of a test particle and hence its velocity in four spherically symmetric solutions for different alternative gravity theories.

\subsection{Motion in Dilaton Coupled Electromagnetic Field}\label{va1}

Static uncharged black hole in general relativity are described by
Schwarzschild solution. If mass of the black hole is much large
compared to Planck mass then this also, to a good approximation,
describes the uncharged black hole in string theory except regions
near singularity. However there was some departure from the
schwarzschild scenario when an exact calculation is made \cite{Yunes}. We shall discuss this solution later in this work.
From now on we shall assume that the above assertion is correct.
However for Einstein-Maxwell solutions the string inspired theory
differ widely from the known classical solution i.e. the
Reissner-Nordstr\"{o}m solution.

The dilaton coupling with $F^{2}$ implies that every solution with non zero $F_{\mu \nu} $ will come with a non zero dilaton.
Thus the charged black hole solution in general relativity (which is the Reissner-nordstr\"{o}m solution) appears in a new form in string
theory due to
the presence of dilaton. The effective four dimensional low energy Lagrangian obtained from string theory is,

$$S=\int d^{4}x \sqrt{-g}[-R+e^{-2\Phi}F^{2}+2(\nabla\Phi)^{2}]$$

where $F_{\mu \nu} $ is the Maxwell field associated with a $U(1)$ subgroup of $E_{8}\times E_{8}$ or $Spin(32)/Z_{2}$.
We have set the remaining gauge fields and antisymmetric tensor field $H_{\mu \nu \rho}$ to zero and $\Phi $ is the dilaton field
(\cite{Garfinkle},\cite{Coleman},\cite{Vega},\cite{Bekensteina},\cite{Bekensteinb},\cite{Bocha},\cite{Witten2}).
Extremizing with respect to the $U(1)$ potential $A_{\mu}$, $\Phi$ and $g_{\mu \nu}$ leads to the following field equations,

\begin{eqnarray}\label{n1}
\left.\begin{array}{c}
(a) \nabla _{\mu} \left(e^{-2\Phi}F^{\mu \nu} \right)=0\\
(b) \nabla ^{2}\Phi +\frac{1}{2}e^{-2\Phi}F^{2}=0\\
(c) R_{\mu \nu}=2\nabla _{\mu}\Phi \nabla _{\nu}\Phi +2e^{-2\Phi}F_{\mu \lambda}F^{\lambda}_{\nu}-\frac{1}{2}g_{\mu \nu}e^{-2\Phi}F^{2}
\end{array}\right\}
\end{eqnarray}

The static spherically symmetric solution corresponding to the
above field equation (\ref{n1}) would give the following line
element as, \cite{Garfinkle}

$$ds^{2}=-(1-\frac{2M}{r})dt^{2}+\frac{1}{(1-\frac{2M}{r})}dr^{2}+r(r-e^{2\Phi_{0}}\frac{Q^{2}}{M})d\Omega^{2}$$

where, $d\Omega^{2}=d\theta^{2}+sin^{2}\theta d\phi^{2}$. Once
again due to isometry we have taken our motion in the equatorial
plane such that, $d\Omega^{2}=d\phi^{2}$. Here $\Phi_{0}$ is the
asymptotic value of dilaton and $Q$ represents the black hole
charge. Note that this is almost identical to the Schwarzschild
metric, with a difference that areas of spheres of constant r and
t now depend on Q. In particular the surface
$r=\frac{Q^{2}e^{2\Phi_{0}}}{M}$ is singular. Also $r=2M$ is the
regular event horizon. Also the evolution of the scalar field
$\Phi$ could be given by,

\begin{equation}\label{n2}
e^{-2\Phi}=e^{-2\Phi _{0}}-\frac{Q^{2}}{Mr}
\end{equation}

We can define the dilaton charge as,

$$D=\frac{1}{4\pi}\int d^{2}\sigma^{\mu}\nabla_{\mu}\Phi$$

where the integral is over a two sphere at spatial infinity and $\sigma^{\mu}$ is the normal to the two sphere at spatial infinity.
For charged black hole this leads to,

\begin{equation}\label{n3}
D=-\frac{Q^{2}e^{2\Phi_{0}}}{2M}
\end{equation}

Here D depends on the asymptotic value of dilaton field, which is determined once M and Q are given and is always negative.
Note that the actual dependance on dilaton field is described by, $e^{-\Phi /M_{pl}}$. Since we have walked in the unit $M_{pl}\sim 1$ we have the
term modified to $e^{-\Phi}$. so as $\Phi \rightarrow \Phi _{0}\sim M_{pl}$, this term is expected to become significant.\\

Now we write the above metric in a generalized form,
\begin{equation}\label{v21}
2L=-f(r)\dot{t}^{2}+\frac{1}{f(r)}\dot{r}^{2}+g(r)\dot{\phi}^{2}
\end{equation}

As usual we have $f(r)=(1-\frac{2M}{r})$ and $g(r)=r(r-e^{2\Phi_{0}}\frac{Q^{2}}{M})$, however due to notational simplicity we have taken
them to be simply $f$ and $g$ respectively. Then the velocity has the following expression,

\begin{equation}\label{v22}
v^{2}=\frac{1}{f^{2}}\left( \frac{dr}{dt} \right)^{2}+\frac{g}{f}\left(\frac{d\phi}{dt}\right)^{2}
\end{equation}

The potential has the following expression which could be given by,

\begin{equation}\label{v23}
V^{2}=f\left(1+\frac{L^{2}}{g}\right)
\end{equation}

Thus 4-velocity components are given by for this potential to yield,

\begin{equation}\label{v24}
t^{\mu}=\left(\frac{E}{f},-\sqrt{E^{2}-V^{2}},0,\frac{L}{g}\right)
\end{equation}

Note that for this case as well we have the following result $t^{\mu}t_{\mu}=-1$. If we have used equation ($\ref{v10}$) then we might have obtained
that the velocity has the same expression as that given by ($\ref{v22}$).

Also the acceleration has no change only $a^{1}$ is non zero and has the value given by ($\ref{v13}$). If we proceed in an identical way then we
obtain the following result for the velocity of a particle relative to an observer,

\begin{equation}\label{v25}
v^{2}=1-\frac{4}{E^{2}_{1}E^{2}_{2}\left[ \left( \frac{1}{E^{2}_{1}}+\frac{1}{E^{2}_{2}} \right)+
\frac{1}{g}\left( \frac{L_{1}}{E_{1}}-\frac{L_{2}}{E_{2}} \right)^{2} \right]^{2}}
\end{equation}

The most interesting part of this velocity expression corresponds to the fact that when at some finite $r$ the quantity $g=0$ then $v=1$.
So even if it had not go to $r=0$ the particle is seen to move with the velocity of light. However under the same situation as above such that both
the particle and the observer moves with the same impact parameter we obtain that this case is prohibited and the particle has $v$ less than $1$ for
all $r$. This singularity corresponds to $r=e^{2\Phi_{0}}\frac{Q^{2}}{M}$

However this particular result is actually an artifact of our
co-ordinate system. For string theory, the statement that the
spacetime has singularity when $r=e^{2\Phi_{0}}\frac{Q^{2}}{M}$ is
actually irrelevant. Since the strings do not couple to the metric
$g_{\mu \nu}$ but rather to $e^{2\Phi} g_{\mu \nu}$. This metric
appear in string $\sigma$ model. In terms of the string metric the
effective lagrangian would become \cite{Garfinkle},

$$S=\int d^{4}x \sqrt{-g}e^{-2\Phi}\left[-R-4(\nabla \Phi)^{2}+F^{2} \right]$$

Hence the charged black hole metric,

\begin{equation}\label{n4}
ds^{2}_{string}=-\frac{1-2Me^{\Phi _{0}}/\rho}{1-Q^{2}e^{3\Phi _{0}}/M\rho}d\tau ^{2}+\frac{d\rho ^{2}}{\left(1-2Me^{\Phi _{0}}/\rho \right)\left(1-Q^{2}e^{3\Phi _{0}}/M\rho\right)}+\rho ^{2}d\Omega
\end{equation}

This metric is identical to the metric given in equation
$(\ref{i2})$ where we have just rescaled the metric by some
conformal factor which is finite every where outside and on the
horizon. With this choice of metric and the assumption that energy
is high or radius is small we obtain the following expression for
relative velocity,

\begin{equation}\label{v18a}
v^{2}=1-\frac{4}{E^{2}_{1}E^{2}_{2}\left[ \left( \frac{1}{E^{2}_{1}}+\frac{1}{E^{2}_{2}} \right)+
\frac{1}{\rho ^{2}}\left( \frac{L_{1}}{E_{1}}-\frac{L_{2}}{E_{2}} \right)^{2} \right]^{2}}
\end{equation}

This is completely identical to the result in equation ($\ref{v18}$), however the metric is completely different, here the general form would
be $ds^{2}=-\frac{f(r)}{g(r)}dt^{2}+\frac{dr^{2}}{f(r)g(r)}+r^{2}d\Omega ^{2}$ where $f(r)=1-2Me^{\Phi _{0}}/\rho$ and $g(r)=1-Q^{2}e^{3\Phi _{0}}/M\rho$.
Hence we arrive at a very important result that for both the spherically symmetric solution in section (\ref{vs1}) and that for dilaton gravity has
the same velocity profile.

\subsection{Spherically Symmetric Solution in Quadratic Gravity}\label{va2}

In this section we consider a class of alternative theories of gravity in four dimensions defined by modifying the Einstein-Hilbert action through
all possible quadratic, algebraic curvature scalars, multiplied by constants or non-constant couplings as (\cite{Yunes},\cite{Stewart},\cite{Green2}),\\

$S=\int d^{4}x \sqrt{-g}[\kappa R+\alpha _{1}f_{1}(\upsilon)R^{2}+\alpha _{2}f_{2}(\upsilon)R_{ab}R^{ab}+\alpha _{3}f_{3}(\upsilon)R_{abcd}R^{abcd}$

\begin{equation}\label{62}
+\alpha _{4}f_{4}(\upsilon)R_{abcd}^{*}R^{abcd}-\frac{\beta}{2}\left(\nabla _{a}\upsilon \nabla ^{a}\upsilon +2V(\upsilon)\right)+L_{matter}]
\end{equation}

where $g$ is the determinant of the metric $g_{ab}$;
$(R,R_{ab},R_{abcd},R_{abcd}^{*})$ are the Ricci scalar and
tensor, the Riemann tensor and its dual \cite{Yunes2},
respectively; $L_{matter}$ is the lagrangian density for other
matter; $\upsilon$ is a scalar field; $(\alpha _{i},\beta)$ are
coupling constants; and $\kappa=(16\pi G)^{-1}$. All other
quadratic curvature terms are linearly dependent e.g., the Weyl
tensor squared. Theories of this type are motivated from low
energy expansion of string theory (\cite{Deser},\cite{Green}).

Varying equation $(\ref{62})$ with respect to the metric and setting $f_{i}(\upsilon)=1$, we find the modified field equations,

\begin{equation}\label{63}
\kappa G_{ab}+\alpha _{1}H_{ab}+ \alpha _{2}I_{ab}+ \alpha _{3}J_{ab}=\frac{1}{2}T_{ab}^{matter}
\end{equation}

where $T_{ab}^{matter}$ is the stress energy of matter, and,

\begin{eqnarray}\label{64}
\left. \begin{array}{c}
(a) H_{ab}=2R_{ab}R-\frac{1}{2}g_{ab}R^{2}- 2 \nabla _{ab}R+ 2g_{ab}\square R\\
(b) I_{ab}=\square R_{ab}+2R_{abcd}R^{cd}-\frac{1}{2}g_{ab}R_{cd}R^{cd}+\frac{1}{2}g_{ab}\square R -\nabla _{ab}R,\\
(c) J_{ab}=8R^{cd}R_{acbd}-2g_{ab}R^{cd}R_{cd}+4\square R_{ab}-2RR_{ab}+\frac{1}{2}g_{ab}R^{2}-2\nabla _{ab}R
\end{array}\right \}
\end{eqnarray}

with $\nabla _{a}$, $\nabla _{ab}=\nabla _{a}\nabla _{b}$, and $\square = \nabla _{a}\nabla ^{a}$ the first and second order covariant derivative
and the D'Alembertian.
The scalar field equation can be given by,

\begin{equation}\label{64a}
\beta \square \upsilon -\beta \frac{dV}{d\upsilon}=-\alpha _{1}R^{2}-\alpha _{2}R_{ab}R^{ab}-\alpha _{3}R_{abcd}R^{abcd}- \alpha _{4}R_{abcd}^{*}R^{abcd}
\end{equation}\\

The spherically symmetric solution to the above field equations
imposing dynamical arguments could be written using the metric
ansatz as \cite{Yunes},

\begin{equation}\label{65}
ds^{2}=-f_{0}\left[1+\epsilon h_{0}(r)\right]dt^{2}+ f_{0}^{-1}\left[1+\epsilon k_{0}(r)\right]dr^{2}+r^{2}d\Omega ^{2}
\end{equation}

and $\upsilon = \upsilon _{0}+\epsilon \upsilon _{0}$, where
$f_{0}=1-2M_{0}/r$, with $M_{0}$ the bare or GR BH mass and
$d\Omega _{2}$ is the line element on two sphere. The free
functions $(h_{0},k_{0})$ are small deformations about the
Schwarzschild metric.

The scalar field equation can be solved to yield,

\begin{equation}\label{66}
\upsilon _{0}=\frac{\alpha _{3}}{\beta}\frac{2}{M_{0}r}\left(1+\frac{M_{0}}{r}+\frac{4M_{0}^{2}}{3r^{2}} \right)
\end{equation}

We can use this scalar field solution to solve modified field equations to linear in $\epsilon$.
Requiring the metric to be asymptotically flat and regular at $r=2M_{0}$, we find the unique solution $h_{0}=F\left(1+\tilde{h_{0}}\right)$
and $K_{0}=-F\left(1+\tilde{h_{0}}\right)$, where $F=-(49/40)\zeta (M_{0}/r)$ and,

~~~~~~~~~~~~~~~~~~~~~~~~~~~~~~~~~~$\tilde{h_{0}}=\frac{2M_{0}}{r}+\frac{548}{147}\frac{M_{0}^{2}}{r^{2}}+\frac{8}{21}\frac{M_{0}^{3}}{r^{3}}-\frac{416}{147}\frac{M_{0}^{4}}{r^{4}}-\frac{1600}{147}\frac{M_{0}^{5}}{r^{5}}$

\begin{equation}\label{67}
\tilde{k_{0}}=\frac{58}{49}\frac{M_{0}}{r}+\frac{76}{49}\frac{M_{0}^{2}}{r^{2}}-\frac{232}{21}\frac{M_{0}^{3}}{r^{3}}-\frac{3488}{147}\frac{M_{0}^{4}}{r^{4}}-\frac{7360}{147}\frac{M_{0}^{5}}{r^{5}}
\end{equation}

Here we have defined the dimensionless coupling function $\zeta=\frac{\alpha _{3}^{2}}{\beta \kappa M_{0}^{4}}$,
which is of the order of $\epsilon$. Such a solution is most general for all dynamical, algebraic, quadratic gravity theories, in spherical symmetry.
We can define the physical mass $M=M_{0}\left[1+(49/80)\zeta\right]$, such that only modified metric components become $g_{tt}=-f(1+h)$ and
$g_{rr}=f^{-1}(1+k)$ where $h=\zeta /(3f)(M/r)^{3}\tilde{h}$ and $k=-(\zeta / f)(M/r)^{2}\tilde{k}$, and

\begin{equation}\label{68}
\tilde{h}=1+\frac{26M}{r}+\frac{66}{5}\frac{M^{2}}{r^{2}}+\frac{96}{5}\frac{M^{3}}{r^{3}}-\frac{80M^{4}}{r^{4}}
\end{equation}

\begin{equation}\label{69}
\tilde{k}=1+\frac{M}{r}+\frac{52}{3}\frac{M^{2}}{r^{2}}+\frac{2M^{3}}{r^{3}}+ \frac{16M^{4}}{5r^{4}}- \frac{368}{3}\frac{M^{5}}{r^{5}}
\end{equation}

where $f=1-2M/r$. Note from the above expression for metric element that Physical observables are related to renormalized mass $M$ not on bare mass
$M_{0}$.\\


\begin{figure}
\includegraphics[height=3.5in, width=3.5in]{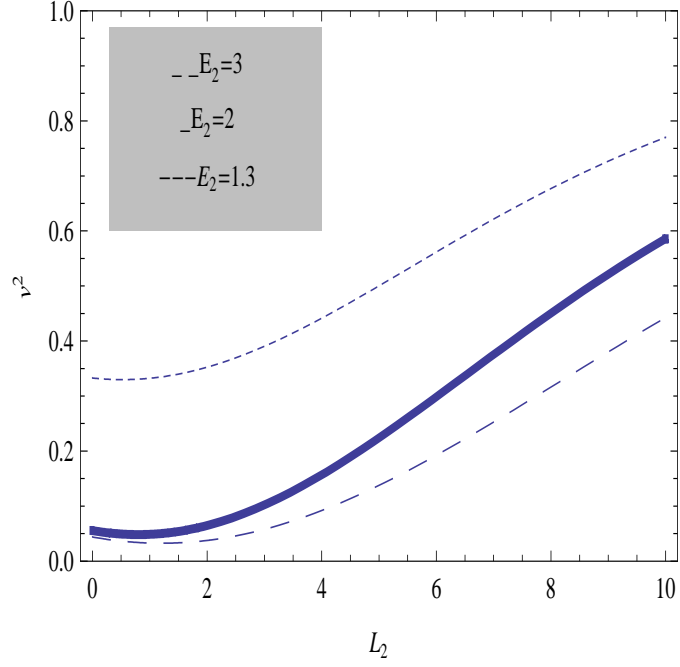}

\caption{The figure shows variation of $v^{2}$ with test particle angular momentum $L_{2}$ for different choices of $E_{2}$.\label{fig5}}

\end{figure}
\begin{figure}
\includegraphics[height=3.5in, width=3.5in]{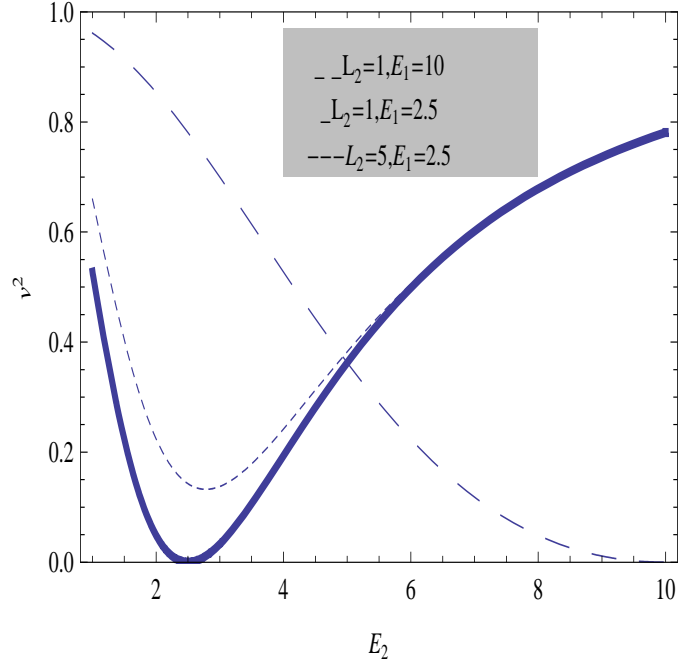}

\caption{The figure shows variation of $v^{2}$ with
test particle energy for different observer energy and test
particle angular momenta.\label{fig6}}

\end{figure}
\begin{figure}
\includegraphics[height=3.5in, width=3.5in]{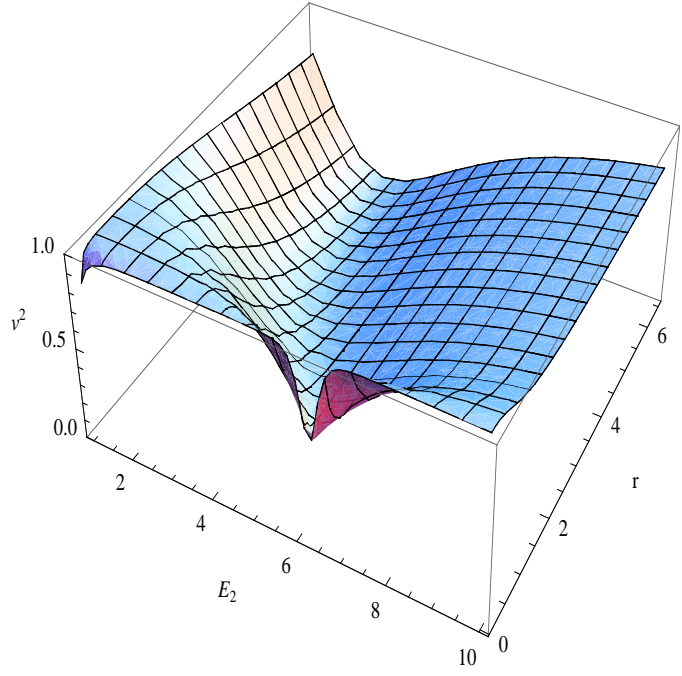}

\caption{The figure shows variation of $v^{2}$ with test particle energy and radial distance.\label{fig7}}

\end{figure}
\begin{figure}
\includegraphics[height=3.5in, width=3.5in]{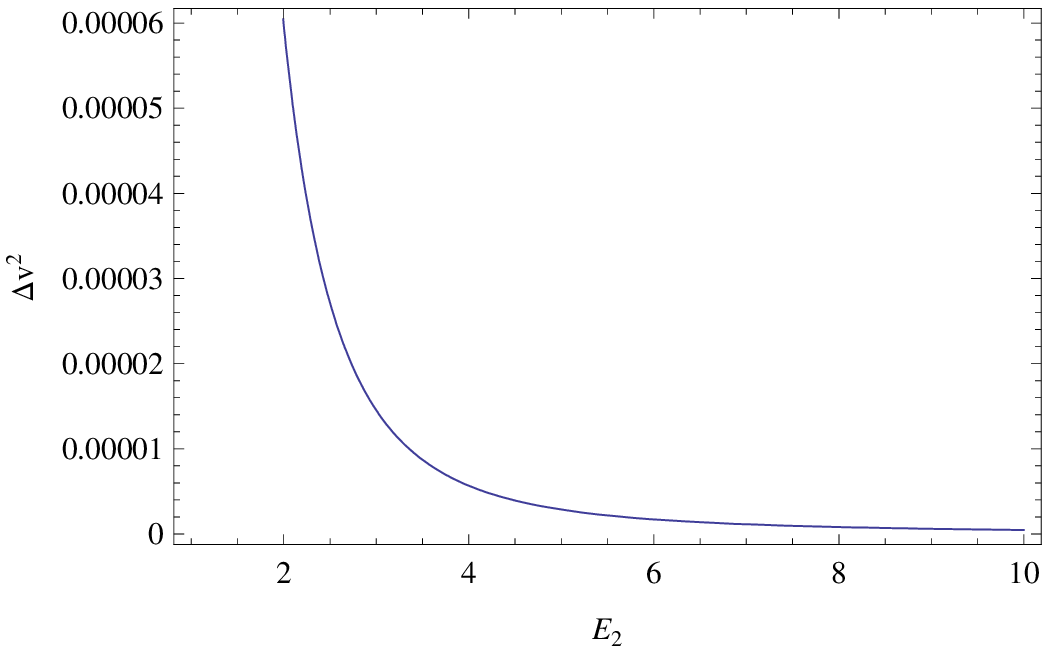}

\caption{The figure shows variation of $\Delta v^{2}$ with $E_{2}$.\label{fig8}}

\end{figure}
\begin{figure}
\includegraphics[height=3.5in, width=3.5in]{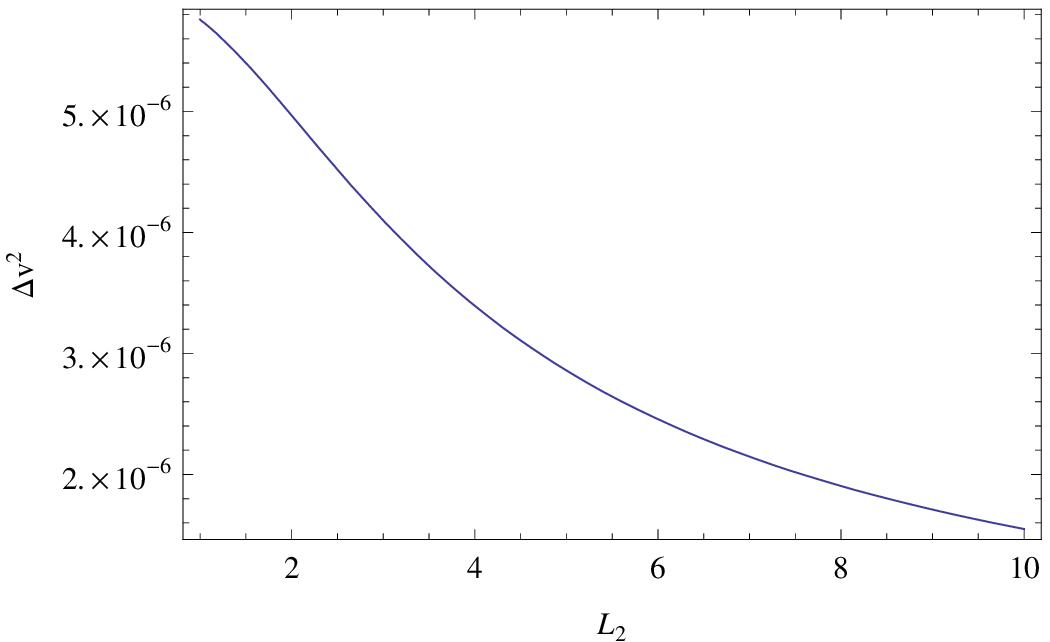}

\caption{The figure shows variation of $\Delta v^{2}$ with $L_{2}$.\label{fig9}}

\end{figure}


In this case the lagrangian has the specific form given by,

\begin{equation}\label{70}
2L=-f(r)\left[1+h(r)\right]\dot{t}^{2}+\frac{\left[1+k(r)\right]}{f(r)}\dot{r}^{2}+r^{2}\dot{\phi}^{2};
\end{equation}

from this we can easily found components of velocity by differentiation. Since the lagrangian does not involve time we have two conserved quantities,
$E$ the energy per particle mass and $L$ the angular momentum per particle mass given by,

\begin{eqnarray}\label{71}
\left.\begin{array}{c}
E=-\frac{\partial L}{\partial \dot{t}}=f(r)\left[1+h(r)\right]\dot{t}\\
L=\frac{\partial L}{\partial \dot{\phi}}=r^{2}\dot{\phi}
\end{array}\right\}
\end{eqnarray}

where the time derivatives are with respect to affine co-ordinate $\tau$. Finally the equation of motion would be given by \citep{Yunes},

\begin{equation}\label{72}
\left(\frac{dr}{d\tau}\right)^{2}=V_{eff}^{GR}-\left[E^{2}h(r)+V_{eff}^{GR}k(r)\right]=V_{eff}
\end{equation}

where we have obtained $V_{eff}^{GR}=E^{2}-f(r)\left[1+\frac{L^{2}}{r^{2}}\right]$. Then the 4-velocity vector could be given by,

\begin{equation}\label{73}
t^{\mu}=\left(\frac{E}{f(1+h)},\sqrt{V_{eff}},0,\frac{L}{r^{2}}\right)
\end{equation}

we can easily check that $t^{\mu}t_{\mu}=-1$. Now we can proceed in an identical way as presented in the previous two sections and that finally
leads to the following expression for relative velocity of a particle with respect to an observer in this space-time to yield,

\begin{eqnarray}\label{v71}
\begin{array}{c}
v^{2}=v^{2}_{GR}-\Delta v^{2}=1-\frac{4}{E^{2}_{1}E^{2}_{2}\left[ \left( \frac{1}{E^{2}_{1}}+\frac{1}{E^{2}_{2}} \right)+
\frac{1}{\rho ^{2}}\left( \frac{L_{1}}{E_{1}}-\frac{L_{2}}{E_{2}} \right)^{2} \right]^{2}}
-\frac{2f^{2}}{E^{2}_{1}E^{2}_{2}\left[1-\frac{fL_{1}L_{2}}{E_{1}E_{2}r^{2}}- \sqrt{E^{2}_{1}-
V^{2}_{1}}\sqrt{E^{2}_{2}-V^{2}_{2}}\right]^{3}} \\
\left[h\left(1-\frac{fL_{1}L_{2}}{E_{1}E_{2}r^{2}}-
\sqrt{E^{2}_{1}-V^{2}_{1}}\sqrt{E^{2}_{2}-V^{2}_{2}}\right) + 2 \left(\frac{V_{1}V_{2}}
{E_{1}E_{2}}\left(h+k+\frac{1}{2}\left(E_{1}\frac{\delta V_{1}}{V^{2}_{1}}+E_{2}\frac{\delta V_{2}}
{V^{2}_{2}}\right)\right)+\frac{L_{1}L_{2}fh}{r^{2}E_{1}E_{2}}\right)\right]
\end{array}
\end{eqnarray}

where we have defined $V_{1}=V_{eff}^{GR}(E_{1},L_{1})$ and
similarly $V_{2}=V_{eff}^{GR}(E_{2},L_{2})$ with similar
interpretation such that $\delta V=-hE^{2}-kV_{eff}^{GR}$. Here
the quantities $f,h,k$ are defined earlier, among them
$f(r)=1-2M/r$ and $h,k$ are given by equations (\ref{68}) and
(\ref{69}). Also note that first two terms are just the velocity
expression we have obtained in equation (\ref{v18}) for a general
spherically symmetric solution and in equation (\ref{v18a}) for
dilaton coupled gravity and refereed to $v_{GR}^{2}$. Also note
that the last term which is the correction term due to alternative
gravity has a negative contribution and when $\zeta =0$ then we
recover our original equation (\ref{v18}).

Figure-$\ref{fig5}$ and figure-$\ref{fig6}$ represents the variation of $v^{2}$ with
test particle angular momentum and energy respectively, as well as
figure-$\ref{fig7}$ represents the variation with both test particle energy
and radial distance. We can very easily verify by comparison with
previous graphs that the effect of introducing quadratic terms in
the action alters the velocity profile near $r=0$ and for low test
particle energy and angular momentum. The effect of test particle
energy and angular momentum on the extra piece $\delta v^{2}$ is
shown in the figure-$\ref{fig8}$ and figure-$\ref{fig9}$, which verifies our
previous assertion. At low energy and angular momentum the
velocity is mostly dictated by the gravitational effect of the
source and that is when the effect of introduction of quadratic
terms could be evident. Hence the above result can be interpreted
as a astrophysical manifestation of the stringy signature, as
these quadratic terms come from some high energy effective string
theory.
\subsection{Motion in Einstein-Maxwell-Gauss-Bonnet Gravity}\label{va3}

Theories with extra spatial dimension have been an active area of
interest even since the original work of Kaluza and Klein, and the
advent of string theory which predicts the presence of extra
spatial dimension. Among many alternatives the Brane world
scenario is considered as a strong candidate which has theoretical
basis in some underlying string theory. Usually, the effect of
string theory on classical gravitational physics (\cite{Green2},\cite{Davies}) is investigated by means of a low
energy effective action, which in addition to the Einstein-Hilbert
action contain squares and higher powers of curvature term.
However the field equations become fourth order and brings in
ghosts \cite{Zumino}. In this context Lovelock \cite{Lovelock} showed that if the higher curvature terms appear
in a particular combination, the field equation become second
order and consequently the ghosts disappear.

In Einstein-Maxwell-Gauss-Bonnet (EMGB) gravity, the action in five dimensional spacetime ($M,g_{\mu \nu}$) can be written as,

\begin{equation}\label{46}
S=\frac{1}{2}\int _{M} d^{5}x \sqrt{-g} \left[R+\alpha L_{GB}+L_{matter} \right],
\end{equation}

where $L_{GB}=R_{\alpha \beta \gamma \delta}R^{\alpha \beta \gamma
\delta}-4R_{\mu \nu}R^{\mu \nu}+R^{2}$ is the GB Lagrangian and
$L_{matter}=F^{\mu \nu}F_{\mu \nu}$ is the Lagrangian for the
electromagnetic field. Here $\alpha$ is the coupling constant of
the GB term having dimension $(length)^{2}$. As $\alpha$ is
regarded as inverse string tension, so $\alpha \geq 0$.

The gravitational and electromagnetic field equations obtained by varying the above action with respect to $g_{\mu \nu}$ and $A_{\mu}$
we could have obtained (see \cite{Chakraborty}),

\begin{eqnarray}\label{47}
\left. \begin{array}{c}
G_{\mu \nu}-\alpha H_{\mu \nu}=T_{\mu \nu}\\
\bigtriangledown _{\mu}F^{\mu}_{\nu}=0\\
H_{\mu \nu}=2\left[RR_{\mu \nu}-2R_{\mu \lambda}R^{\lambda}_{\mu}-2R^{\gamma \delta}R_{\mu \gamma \nu \delta}+R^{\alpha \beta \gamma}_{\mu}R_{\nu \alpha \beta \gamma} \right]-\frac{1}{2}g_{\mu \nu}L_{GB}
\end{array}\right \}
\end{eqnarray}

where $T_{\mu \nu}=2F^{\lambda}_{\mu}F_{\lambda \nu}-\frac{1}{2}F_{\lambda \sigma}F^{\lambda \sigma}g_{\mu \nu}$ is the electromagnetic field tensor.

A spherically symmetric solution to the above action has been obtained by \cite{Dehghani} and the line element is given by,

\begin{equation}\label{48}
ds^{2}=-g(r)dt^{2}+\frac{dr^{2}}{g(r)}+r^{2}d\Omega _{3}^{2},
\end{equation}

where the metric co-efficient is,

\begin{equation}\label{49}
g(r)=K+\frac{r^{2}}{4\alpha}\left[1\pm \sqrt{1+\frac{8\alpha \left(m+2\alpha \mid K \mid \right) }{r^{4}} -\frac{8\alpha q^{2}}{3r^{6}}} \right]
\end{equation}

Here $K$ is the curvature, $m+2\alpha \mid K\mid$ is the geometrical mass and $d\Omega _{3}^{2}$ is the metric of a 3D hypersurface such that,

\begin{equation}\label{50}
d\Omega _{3}^{2}=d\theta _{1}^{2}+sin^{2}\theta _{1}\left( d\theta _{2}^{2}+sin^{2}\theta _{2}d\theta _{3}^{2}\right)
\end{equation}

The range is given by $\theta _{1},\theta _{2}:[0,\pi]$.
We assume that there is a constant charge $q$ at $r=0$ and the vector potential be $A_{\mu}=\Phi (r)\delta_{\mu}^{0}$ such that
$\Phi (r)=-\frac{q}{2r^{2}}$.

In this metric the metric function $g(r)$ will be real for $r \geq r_{0}$ where $r_{0}^{2}$ is the largest real solution of the cubic equation,

\begin{equation}\label{51}
3z^{3}+24\alpha \left(m+2\alpha \mid K \mid \right)z-8\alpha q^{2}=0
\end{equation}

By a transformation of the radial co-ordinates we can show that $r=r_{0}$ is an essential singularity of the spacetime.
We shall choose $K=1$ and shall consider the $-$ve sign in front of square root of equation $(\ref{49})$ which leads to asymptotically flat solution.

However note that the line element as presented in equation ($\ref{48}$) is exactly of the same form as we have used in equation ($\ref{i1}$). Thus the velocity of a test particle relative to an observer would have the same form as presented in equation ($\ref{v18}$). Thus all the properties of this velocity remain valid in this EMGB gravity and shows the usefulness of our definition of velocity.

\subsection{Motion in F(R) gravity}\label{va4}

General Relativity (GR) is a widely accepted as a fundamental
theory relating matter energy density to geometric properties of
spacetime. The standard big-bang cosmological model can explain
the evolution of the universe well except inflation and late time
cosmic acceleration. Although many scalar field models have been
constructed in the frame work of string theory and supergravity to
explain inflation but Cosmic Microwave Background radiation still
do not show any evidence in favor of a particular model. The same
kind of approach is also taken to explain cosmic acceleration by
introducing different dark energy models where also concrete
observation is still lacking.

Thus one of the simplest choice is to modify GR action by introducing a term $F(R)$
in the lagrangian, where $F$ is an arbitrary function of scalar curvature $R$. There exists
two methods for deriving field equations, first, by varying the action with respect to metric tensor $g_{\mu \nu}$. The other method called
Palatini method should not be discussed here .In F(R) gravity (\cite{nel10},\cite{cor10},\cite{bal10},\cite{fel10}), the scalar curvature $R$
in the Einstein-Hilbert action

\begin{equation}\label{va11}
S_{EH}=\int d^{4}x \sqrt{-g}\left(\frac{R}{16\pi} +L_{matter}\right),
\end{equation}

gets replaced by an appropriate function of scalar curvature:

\begin{equation}\label{va12}
S_{F(R)}=\int d^{4}x \sqrt{-g}\left(\frac{F(R)}{16\pi} +L_{matter}\right)
\end{equation}

Varying this action we readily obtain the corresponding field equation to be given by,

\begin{equation}\label{va13}
\frac{1}{2}g_{\mu \nu}F(R)-R_{\mu \nu}F'(R)-g_{\mu \nu}\square F'(R)+\nabla _{\mu}\nabla _{\nu}F'(R) =-4\pi T_{matter \mu \nu}
\end{equation}

Several solutions (often exact) to this field equation may be found but due to complicated nature of field equations the number of such exact
solutions are much less than that in general relativity. Without any matter and assuming the Ricci tensor to be covariantly
constant equation ($\ref{va13}$) reduces to the following algebraic equation,

\begin{equation}\label{va14}
0=2F(R)-RF'(R)
\end{equation}

From the above equation we can show that Schwarzchild-(anti-)de
Sitter space is an exact vacuum solution to it. Thus the
respective line element would be given by,

\begin{equation}\label{va15}
ds^{2}=-\left(1-\frac{2M}{r}\mp \frac{r^{2}}{L^{2}}\right)dt^{2}+ \left(1-\frac{2M}{r}\mp \frac{r^{2}}{L^{2}}\right)^{-1}dr^{2}+r^{2}d\Omega ^{2}
\end{equation}

Here the minus and plus sign corresponds to de Sitter and anti de Sitter space respectively, $M$ is the mass of the black hole
and $L$ is the length parameter of (anti-)de Sitter space, which is related to the curvature $R=\pm \frac{12}{L^{2}}$ (the plus sign corresponds
to de Sitter space and minus sign corresponds to anti de Sitter space).

The vacuum solution for $F(R)$ gravity also has the same form as
we have used in equation ($\ref{i1}$). Thus all the results of
section \label{vs1} will remain valid here as well. Hence the
relative velocity will have the same characteristics in vacuum
solution for $F(R)$ gravity theory as well. This justifies our
assertion as stated in section $\ref{vc}$.

\section{Discussion}\label{vd}

We have shown that velocity of any ingoing particle with respect
to observer sets as defined in the section ($\ref{vc}$) for a
general spherically symmetric potential with unit 2-sphere is
always less than that of light outside the singular point, it
approaches the speed of light as $r\rightarrow 0$. However the
notion of static observers are not valid for $r\leq 2M$. It is
valid only for region outside the event horizon. Thus we have
defined ingoing observers and determine velocity with respect to
the observer. We found that velocity of the test particle always
remain less than 1. For a different choice of metric with a
function on 2-sphere we found that the velocity is always less
than 1 which may not be self-evident in one set of co-ordinates,
but by going to another set we have actually shown that the
previous results are retained. Finally the spherically symmetric
solution in quadratic gravity shows another instance of the
correctness of our result. However there we have obtained a
correction factor to the velocity expression due to presence of
quadratic terms and hence this directly shows that the velocity
profile of an object differ considerably in alternative theories
from the result in Einstein gravity. However that particular
correction term would be Planck suppressed and hence very
difficult to observe, however just out side the event horizon of
the BH, where the tidal effects are huge these effects can in
principle be observed. For the other two theories we
have obtained the same expression as for the general spherically
symmetric model. Thus they follow our previous assertion
connecting to the relative velocity of a test particle. Also it
should be noted that the above analysis is not restricted to
Einstein gravity or the solutions we have discussed, it can also
be applied to other spherically symmetric black hole solutions in
other modified gravity theories. Also it could be extended to
higher dimensional black holes.
Extension to rotating black holes would be an interesting work for the future.\\

\acknowledgements The author thanks prof. Subenoy
Chakraborty of Jadavpur University and Prof. Soumitra Sengupta of
IACS for helpful discussion. The author also thanks DST, Govt. of
India for awarding KVPY fellowship. He gratefully thanks IUCAA,
Pune, for warm hospitality where a part of this work was done.


\end{document}